\documentclass{PoS}
\input{epsf} 

\newcommand\as{\alpha_{\mathrm{S}}}

\title{NNLO QCD correction to vector boson production at hadron colliders}

\ShortTitle{NNLO QCD correction to vector boson production at hadron colliders}

\author{\speaker{Giancarlo Ferrera}\\%
        Dipartimento di Fisica, Universit\`a di Firenze and I.N.F.N. Sezione di Firenze\\
        I-50019 Sesto Fiorentino, Florence, Italy\\
        E-mail: \email{ferrera@fi.infn.it}}

\abstract{
We present a fully-exclusive next-to-next-to-leading order (NNLO) QCD calculation for 
vector boson production in hadron-hadron collisions.
The calculation is implemented in a parton level Monte Carlo program, which includes
$\gamma-Z$ interference, finite-width effects, the leptonic decay of the vector bosons and
the corresponding spin correlations. 
The code allows
the user to apply arbitrary (though infrared safe) kinematical cuts on the final-states and to 
compute distributions in the form of bin histograms.
We show some illustrative numerical results at the Tevatron and the LHC.
}

\FullConference{RADCOR 2009 - 9th International Symposium on Radiative Corrections (Applications of Quantum Field Theory to Phenomenology) ,\\
		 October 25 - 30 2009\\
		 Ascona, Switzerland}

\begin{document}

Vector boson production in hadron collisions, 
the well known Drell-Yan (DY) process\cite{Drell:1970wh}, has
a special role for physics studies at hadron colliders. 
Having large production rates with relatively simple experimental signatures, 
this process is important for detectors calibration, it gives stringent
informations on Parton Distribution Functions (PDFs), is a strong test for perturbative QCD
predictions and may signals effects from physics beyond the Standard Model.

It is therefore essential to have accurate theoretical predictions for the vector-boson production
cross sections and related distributions, which are predicted by perturbative QCD as an expansion
in the strong coupling $\alpha_S$. 
The next-to-next-to-leading order (NNLO) QCD corrections (i.e. $\mathcal{O}(\alpha_S^2)$)
have been calculated analytically for the total cross section~\cite{Hamberg:1990np} 
and the rapidity distribution of the vector
boson~\cite{Anastasiou:2003ds}. The fully exclusive NNLO calculation
has also been performed~\cite{DYdiff,DYNNLO}.
Furthermore, electroweak corrections up to $\mathcal{O}(\alpha)$ have been computed for both $W$~\cite{ewW} 
and $Z$ production~\cite{ewZ}.

The computation of higher-order QCD corrections 
to hard-scattering processes is a hard task. 
Difficulties arise from 
the presence
of infrared  singularities at intermediate stages of the calculation that prevent
a straightforward implementation of numerical techniques.
For the above reason, fully exclusive cross-sections in hadron collisions have been computed so far only for
Higgs boson production~\cite{Anastasiou:2003gr,Hdiff,Catani:2007vq,Grazzini:2008tf} 
and the Drell--Yan process~\cite{DYdiff,DYNNLO}.

In this contribution
we present a recent fully exclusive NNLO QCD calculation 
for vector boson production in hadron collisions~\cite{DYNNLO}.
The calculation uses the NNLO extension of subtraction formalism 
introduced in Ref.~\cite{Catani:2007vq}. The method 
is valid in general for the production of colourless high-mass systems in hadron collisions.

%%%%%%%%%%%%%%%%%%%%%

We consider the hard-scattering process:
\begin{equation}
h_1+h_2\to V(q)+X,
\end{equation}
where  the colliding hadrons $h_1$ and $h_2$ produce
the vector boson $V$ ($V=Z/\gamma^*, W^+$ or $W^-$), 
with four-momentum $q$ and high invariant mass $\sqrt {q^2}$,
plus  an inclusive final state $X$.

Following Ref.~\cite{Catani:2007vq}, we observe that, at LO, 
the transverse momentum $q_T$ of $V$ is exactly zero.
This means that, as long as $q_T\neq 0$,
the (N)NLO contributions are given by the (N)LO contributions to the final state
$V + jet(s)$~\cite{Giele:1993dj}:
\begin{equation}
%$\,
d{\hat\sigma}^{V}_{(N)NLO}|_{q_T\neq 0}\,=\,d{\hat\sigma}^{V+{\rm jets}}_{(N)LO}\,\,.
\end{equation}
%$.
We compute $d{\hat\sigma}^{V+{\rm jets}}_{NLO}$ by using the 
subtraction method at NLO~\cite{Frixione:1995ms,Catani:1996vz}  and
we treat the remaining NNLO singularities at $q_T = 0$ by the additional 
subtraction of an 
universal~\footnote{It depends only on the flavour of the initial-state partons involved in the LO partonic subprocess.} 
counter-term $d{\hat\sigma}^{CT}_{(N)LO}$ %
constructed by exploiting the universality of the
logarithmically-enhanced contributions to the transverse momentum distribution%
~\footnote{For the explicit
form of the counter-term 
see Refs.\cite{Catani:2007vq,hqt}.}.
Schematically we have 
\begin{equation}
\label{main}
d{\hat\sigma}^{V}_{(N)NLO}=
{\cal H}^{V}_{(N)NLO}\otimes d{\hat\sigma}^{V}_{LO}
+\left[ d{\hat\sigma}^{V+{\rm jets}}_{(N)LO}-
d{\hat\sigma}^{CT}_{(N)LO}\right]\;\; ,
\end{equation}
where $\mathcal{H}^V_{(N)NLO}$ is a process-dependent coefficient function 
necessary to reproduce the correct normalization~\cite{deFlorian:2000pr,DYNNLO}.

    We have encoded our NNLO computation in a parton level 
Monte Carlo event generator. 
The calculation includes finite-width effects,
the $\gamma-Z$ interference,
the leptonic decay of the vector bosons and the corresponding 
spin correlations. 
Our numerical code is particularly suitable
for the computation 
of distributions in the form of bin histograms.

%%%%%%%%%%%%%%%%%%%%%

In the following
we present some illustrative
numerical results for $Z$ and $W$ production at the Tevatron and the LHC.
We consider $u,d,s,c,b$ quarks in the initial state.
In the case of
$W^\pm$ production,
we use the (unitarity constrained) CKM matrix elements $V_{ud}=0.97419$, $V_{us}=0.2257$, $V_{ub}=0.00359$,
$V_{cd}=0.2256$, $V_{cs}=0.97334$, $V_{cb}=0.0415$ from the PDG 2008 \cite{Amsler:2008zzb}. 
We use the so called $G_\mu$ scheme for the electroweak couplings,
with the following input parameters: $G_F = 1.16637\times 10^{-5}$~GeV$^{-2}$,
$m_Z = 91.1876$~GeV, $\Gamma_Z=2.4952$~GeV, $m_W = 80.398$~GeV
and $\Gamma_W=2.141$~GeV.
As for the PDFs, we use 
MSTW2008 \cite{Martin:2009iq} as default set,
 evaluating $\as$ at each corresponding order
(i.e., we use $(n+1)$-loop $\as$ at N$^n$LO, with $n=0,1,2$). 
We fix the 
renormalization ($\mu_R$) and factorization ($\mu_F$) scales to the mass of the vector boson $m_V$.

%%====================================
\begin{figure}[htb]
\begin{center}
\begin{tabular}{c}
\epsfxsize=12truecm
\epsffile{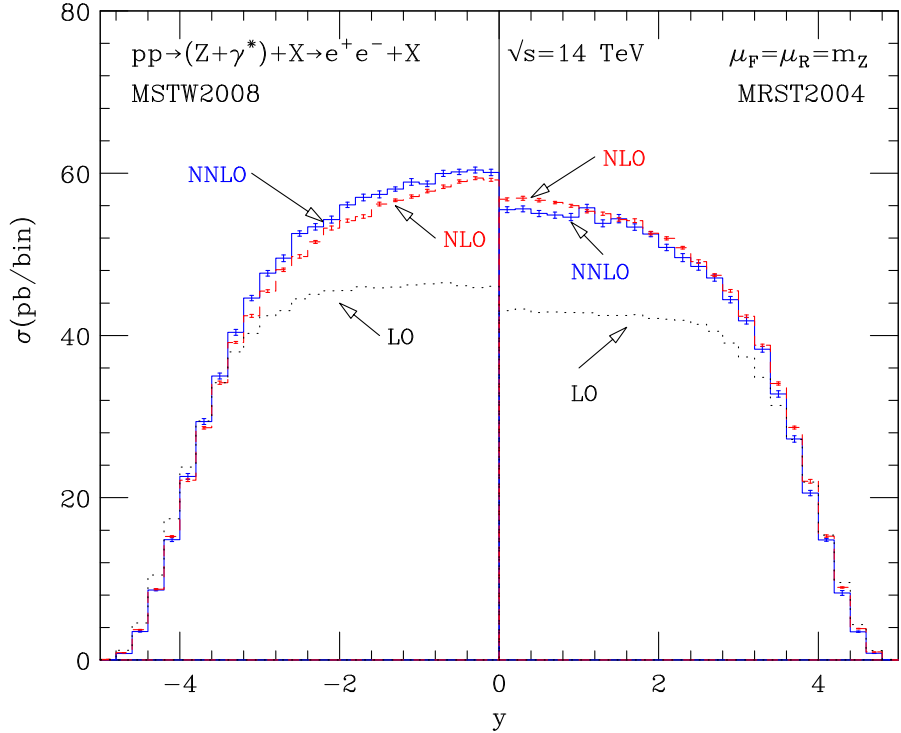}\\
\end{tabular}
\end{center}
\caption{\label{fig:y34lhc}
{\em Rapidity distribution of an on-shell $Z$ boson at the LHC. Results obtained
with the MSTW2008 set (left panel) are compared with those obtained with the MRST2004 set (right panel).}}
\end{figure}
%%====================================
We start the presentation of our results by considering
the inclusive production of $e^+e^-$ pairs from the decay of an on-shell $Z$ boson
at the LHC. In the left panel of Fig.~\ref{fig:y34lhc} we show the rapidity 
distribution of the $e^+e^-$ pair at LO, NLO and NNLO, computed 
by using
the MSTW2008 PDFs~\cite{Martin:2009iq}.
The corresponding cross sections
 are $\sigma_{LO}=1.761 \pm 0.001$~nb, 
$\sigma_{NLO}=2.030 \pm 0.001$~nb and $\sigma_{NNLO}=2.089 \pm 0.003$~nb. 
The total cross section is increased by about $3$\%
in going from NLO to NNLO. In the right panel of Fig.~\ref{fig:y34lhc} 
we show the results obtained by using the 
MRST2002 LO \cite{Martin:2002aw} 
and MRST2004 \cite{Martin:2004ir}
sets of parton distribution functions. The corresponding cross sections are
$\sigma_{LO}=1.629 \pm 0.001$~nb, $\sigma_{NLO}=1.992 \pm 0.001$~nb and 
$\sigma_{NNLO}=1.954 \pm 0.003$~nb. In this case the total cross section 
is decreased by about $2$\%
in going from NLO to NNLO.

%%====================================
\begin{figure}[htb]
\begin{center}
\begin{tabular}{c}
\epsfxsize=13.8truecm
\epsffile{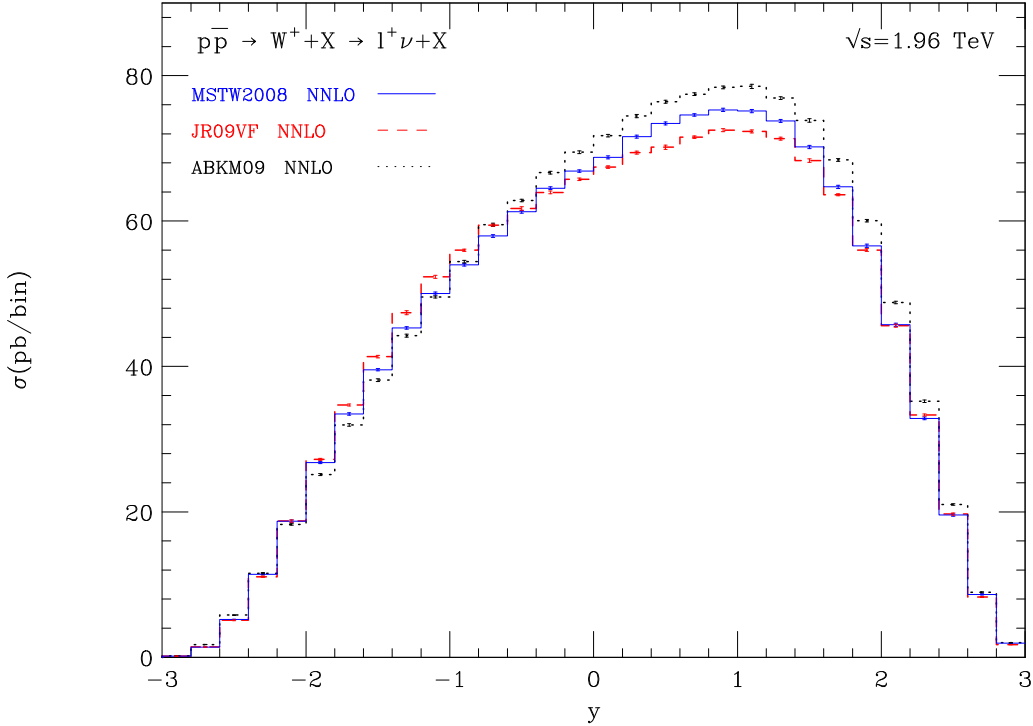}\\
\end{tabular}
\end{center}
\caption{\label{fig:one}
{\em Rapidity distribution of an on-shell $W^+$ boson at the Tevatron. The  NNLO
result obtained
with the MSTW2008 set  are compared with those obtained with the JR09VF 
and ABKM09 sets.}}
\end{figure}
%%====================================
We next consider
the production of an on-shell $W^+$ boson 
at the Tevatron. 
In Fig.~\ref{fig:one} we show the rapidity 
distribution of the $W^+$ at NNLO, computed 
by using
the MSTW2008 PDFs~\cite{Martin:2009iq}. We also show, for comparison, the NNLO prediction by using 
the JR09VF~\cite{JR09VF} and ABKM09~\cite{ABKM} PDFs.
The corresponding total cross sections
 are 
$\sigma_{NNLO}^{(MSTW)}=1.349 \pm 0.002$~nb, $\sigma_{NNLO}^{(JRVF)}=1.338 \pm 0.002$~nb and
$\sigma_{NNLO}^{(ABKM)}=1.391 \pm 0.002$~nb.  The differences between the three results 
can be reach, in the central rapidity region, the level of about 5\%.

%%====================================
\begin{figure}[htb]
\begin{center}
\begin{tabular}{c}
\epsfxsize=12truecm
\epsffile{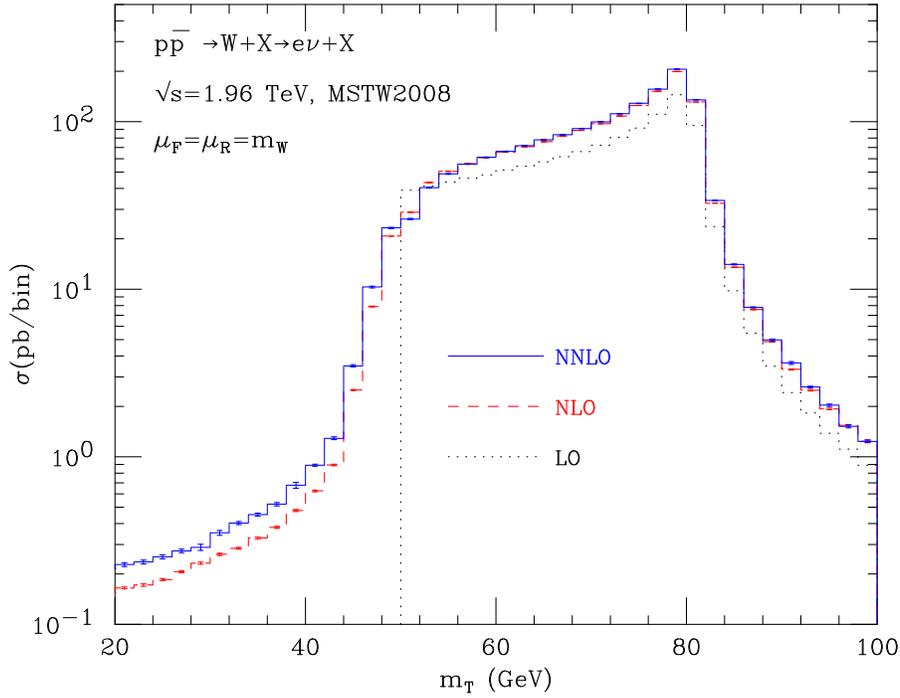}\\
\end{tabular}
\end{center}
\caption{\label{fig:tmass}
{\em Transverse mass distribution for $W$ production at the Tevatron.}}
\end{figure}
%%====================================
We finally consider the production of a charged lepton plus missing $p_T$ 
through the decay of a $W$ boson ($W=W^+,W^-$) at the Tevatron.
The charged lepton is selected to have $p_T>20$~GeV and $|\eta|<2$ and the missing $p_T$
of the event is required to be larger than 25~GeV. 
The transverse mass of the
event is defined as $m_T=\sqrt{2p_T^lp_T^{\rm miss}(1-\cos\phi)}$, where $\phi$ is the angle between the
the $p_T$ of the lepton and the missing $p_T$.
In Fig.~\ref{fig:tmass} we show the transverse mass  distribution at LO, NLO and NNLO:
the accepted cross sections are $\sigma_{LO}=1.161 \pm 0.001$~nb,
$\sigma_{NLO}=1.550\pm 0.001$~nb and $\sigma_{NNLO}=1.586 \pm 0.002$~nb.
Since at LO the $W$ boson is produced with zero 
transverse momentum,
the requirement $p_T^{\rm miss}>25$~GeV sets
$m_T\geq 50$~GeV. As a consequence
at  LO the transverse mass distribution has a kinematical boundary at $m_T=50$~GeV.
Around this boundary 
there are perturbative 
instabilities due to (integrable) logarithmic 
singularities~\cite{Catani:1997xc}.
We also note that, below the boundary, the NNLO corrections to the NLO result
are large.
This is not unexpected, 
since in this region of transverse masses, the ${\cal O}(\as)$ result corresponds
to the calculation at 
the first perturbative order and, therefore, our ${\cal O}(\as^2)$ result 
is actually only a calculation at the NLO level of perturbative accuracy.

%%%%%%%%%%%%

We have presented a fully exclusive NNLO QCD calculation for vector boson production
in hadron-hadron collisions. 
Our calculation is
directly implemented in a parton level event generator.
This feature makes it particularly suitable for practical applications
to the computation of distributions in the form of bin histograms.
For illustrative purpose, we have shown some selected numerical distributions
at the Tevatron and the LHC.

\end{document}